\documentclass{optica-article}

\journal{opticajournal} 

\articletype{Research Article}
\usepackage{lineno}
\begin{document}
\title{Terawatt-level three-stage pulse compression for all-attosecond pump-probe spectroscopy} 

\author{Eli Sobolev, Mikhail Volkov, Evaldas Svirplys, John Thomas, Tobias Witting, Marc J. J. Vrakking, Bernd Schütte\authormark{*}}
\address{Max-Born-Institut, Max-Born-Strasse 2A, 12489 Berlin, Germany}
\email{\authormark{*}Bernd.Schuette@mbi-berlin.de}

\begin{abstract*}
The generation of terawatt (TW) near-single-cycle laser pulses is of high interest for applications including attosecond science. Here we demonstrate a three-stage post-compression scheme in a non-guided geometry using He as the nonlinear medium, resulting in the generation of multi-mJ pulses with a duration of 3.7~fs. Key features of this approach are its  simplicity, robustness and high stability, making it ideally suited for highly demanding applications such as attosecond-pump attosecond-probe spectroscopy (APAPS). This is demonstrated by performing two-color APAPS in Ar and Ne, where both simultaneous and sequential two-photon absorption are observed. Our approach is scalable to multi-TW powers.
\end{abstract*}

\section{Introduction}
One of the longstanding ambitions in the field of attosecond science is the development of attosecond-pump attosecond probe-spectroscopy (APAPS). This application of attosecond laser pulses requires the availability of intense attosecond pulses (typically $> 10^{13}$~W/cm$^2$), since a sufficiently high probability needs to be created so that the sample under investigation will absorb a photon from both the attosecond pump and the attosecond probe laser. Thus far, the number of APAPS experiments in the scientific literature is limited, due to the low efficiency of the high-harmonic generation (HHG) technique that is most commonly used for generating attosecond laser pulses, where the conversion efficiency from the driving laser into the extreme-ultraviolet (XUV) or soft X-ray regime is usually $<10^{-4}$. Accordingly, previous HHG-based demonstrations of APAPS have been sparse and have required the use of very high pulse energy, low repetition rate ($10-100$~Hz) laser systems~\cite{tzallas11, takahashi13, kretschmar22}. Moreover, they were limited to measurements of relatively easily obtainable observables (ion spectroscopy). As a result of this slow progress, many researchers have turned towards free-electron lasers in recent years as a means to pursue APAPS experiments. The generation of intense isolated attosecond laser pulses was demonstrated at the Linac Coherent Light Source (LCLS)~\cite{duris20} and has led to first attosecond time-resolved measurements using a two-color X-ray and near-infrared (NIR) pump-probe scheme~\cite{li22}. Very recently, first demonstrations of APAPS were published~\cite{guo24, li24}. These results were, however, limited to only one~\cite{li24} or four time delays~\cite{guo24}.

It remains a highly desirable objective to further develop HHG-based APAPS, permitting the implementation of more sophisticated measurement schemes such as all-attosecond transient absorption spectroscopy (AATAS). To this end, a simple, robust and stable scheme is required for the generation of terawatt (TW)-level near-single-cycle laser pulses at kHz repetition rates that can be used as driver lasers for the generation of isolated attosecond laser pulses. In the last few decades, several approaches have been applied towards the generation of intense few-cycle laser pulses. Following a first demonstration by Nisoli \textit{et al.}~\cite{nisoli97}, post-compression in long, gas-filled hollow-core fibers (HCFs) based on self-phase modulation (SPM) has been extensively pursued, meanwhile resulting in $\leq 5$~fs pulses with energies ranging from 3.5 to 6~mJ at kHz repetition rates~\cite{bohman10, quille20, nagy20}. Alternatively, the generation of 3.9-fs pulses with a peak power $>2$~TW was reported, based on spectral broadening in a thin fused silica plate~\cite{toth23}. Furthermore, multi-TW few-cycle laser pulses were generated using waveform synthesizers without the need for post-compression~\cite{rivas17, xue20}. While the energy scaling of HCFs is quite sophisticated~\cite{nagy21}, the approaches reported in Refs.~\cite{toth23, rivas17, xue20} relied on the development of highly specialized laser systems and were demonstrated at a repetition rate of only 10~Hz, which is a limiting factor for many applications that require a high stability and good statistics. Thus, although the afore-mentioned works demonstrate that the generation of TW-level near-single-cycle pulses has been possible by means of a number of techniques, to the best of our knowledge, no APAPS results have yet been reported using any of these laser systems. 

In addition to the afore-mentioned methods, several alternative post-compression schemes have successfully been developed, but have so far not resulted in the generation of TW near-single-cycle pulses. Among these, multipass cells enable the compression of pulses with high pulse energies, but the generation of near-single-cycle pulses remains challenging~\cite{viotti22}. Compression via filamentation using one~\cite{steingrube12} or two stages~\cite{hauri04} has led to the generation of few-cycle pulses, but has so far been limited to pulse energies below 1~mJ. Recently, Tsai \textit{et al.} demonstrated a cascaded four-stage compression scheme at a pulse energy of $\approx1$~mJ based on SPM in Ar~\cite{tsai22}. This approach resulted in an excellent pulse energy stability, allowing them to perform attosecond streaking. In their scheme, Tsai \textit{et al.} operated at a relatively long driving laser wavelength of 1030~nm Ref.~\cite{tsai22}, which is very well suited for the generation of high XUV and soft X-ray photon energies, at the cost of a reduced HHG conversion efficiency. The latter was experimentally shown to scale as $\lambda_{\textrm{driver}}^{-6.5}$~\cite{shiner09} for HHG in Kr, where $\lambda_{\textrm{driver}}$ refers to the driver laser wavelength. Correspondingly, the HHG conversion efficiency may be expected to be lower by a factor of about 5 when compared to a driving laser wavelength of 800~nm, limiting the suitability of this source for APAPS.

\begin{figure}[tb]
 \centering
 \includegraphics[width=0.5\textwidth] {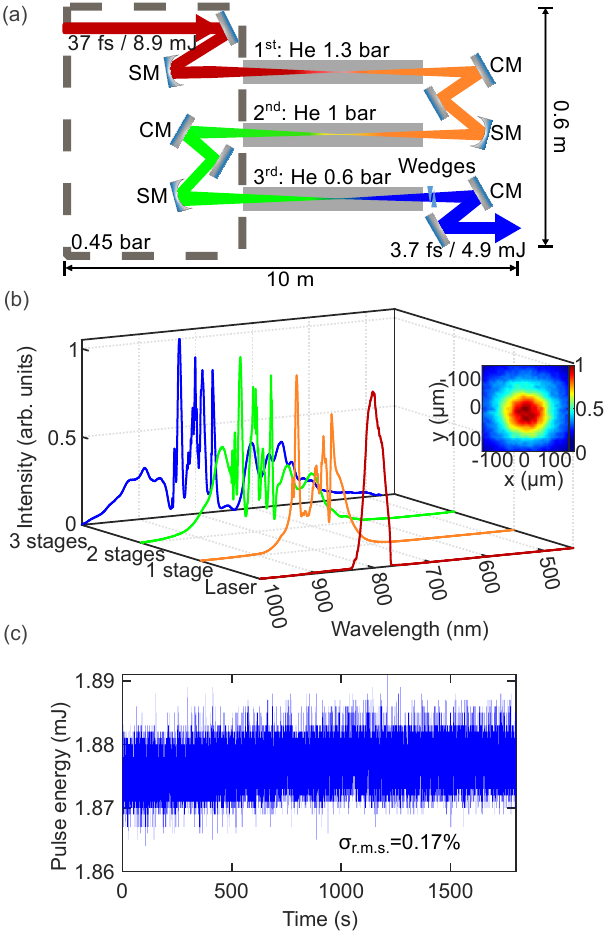}
 \caption{\label{figure1} Three-stage pulse compression. (a) Scheme of the compression setup, consisting of three tubes filled with He at pressures of 1.3~bar, 1.0~bar and 0.6~bar, and a chirped-mirror (CM) compressor after each stage. A single spherical mirror (SM) is used to focus or image the pulses before each stage. The dispersion is fine-controlled before the third stage by adjusting the air pressure within a vacuum chamber, using an optimal value of 0.45~bar. After the third stage, two wedges are used to fine-control the chirp. (b) NIR spectra obtained from the laser (red), and after each of the three compression stages (orange, green and blue). The NIR profile measured in the focus used for HHG is shown in the inset. (c) Single-shot NIR pulse energy stability after the final compression stage, where an r.m.s. value of 0.17~$\%$ was obtained over half an hour.}
\end{figure}

In our laboratory, we have extended the cascaded compression technique introduced by Tsai \textit{et al.} to an 800~nm central wavelength and to higher pulse energies. To be able to increase the pulse energy that can be temporally compressed, (i) a loose-focusing geometry is adopted that maximizes the interaction length while avoiding a high degree of ionization, and (ii) helium is chosen as the interaction medium, since it has the highest ionization potential of all elements. The approach benefits from a high degree of stability and robustness, making it ideally suited for demanding applications such as APAPS. As a benchmark, we present high signal-to-noise APAPS experiments that are based on use of the compressed laser pulses. 

\begin{figure}[tb]
 \centering
 \includegraphics[width=\textwidth] {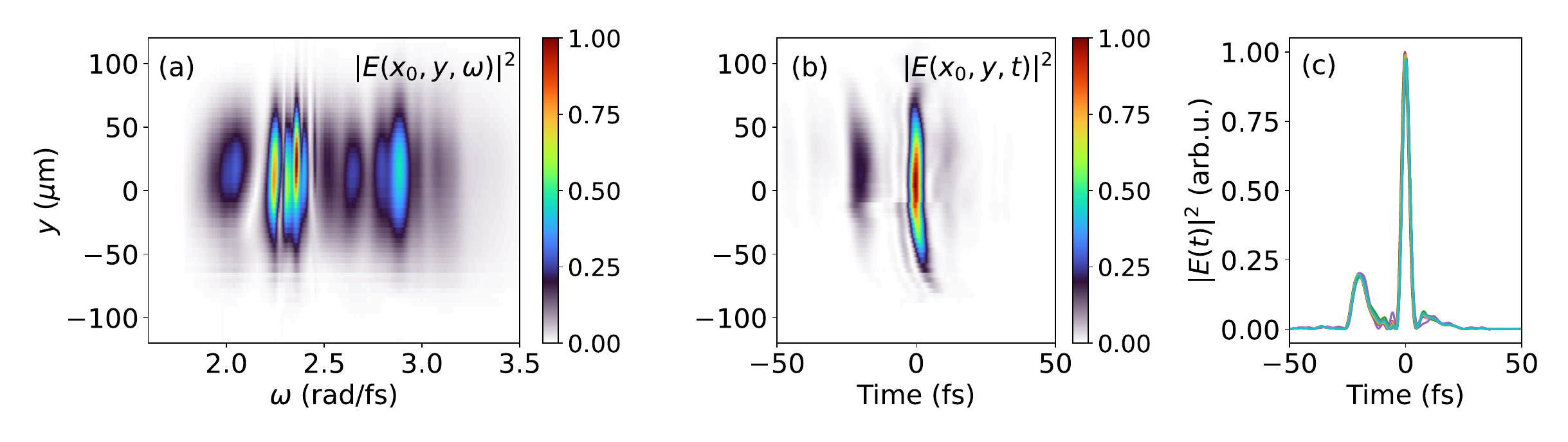}
 \caption{\label{figure2} Temporal characterization. (a) Spatio-spectral intensity distribution $|E(x_0, y, \omega)|^2$ and (b) spatio-temporal intensity distribution $|E(x_0, y, t)|^2$. (c) The integrated temporal intensity profiles $|E(t)|^2$ are shown for ten sequential measurements taken over the time-span of 1 minute. The measured pulse duration of $3.69 \pm 0.04$~fs corresponds to $1.54 \pm 0.02$ optical cycles.}
\end{figure}

\section {Temporal compression}

Temporal compression was performed for the output of a Ti:sapphire laser system (Spitfire Ace, Spectra Physics). Near-infrared (NIR) pulses with a pulse energy of 8.9~mJ and a duration of 37~fs at 1~kHz were used. As shown in Fig.~\ref{figure1}(a), the NIR pulses were focused into a first tube (filled with He at a pressure of 1.3~bar) using a spherical mirror with a focal length of 5~m. This was followed by temporal compression using four chirped mirrors (PC125, Ultrafast Innovations), with a group delay dispersion (GDD) of $-60$~fs$^2$ each. A spherical mirror with a focal length of 2.5~m was used to image the pulses into a second tube (filled with He at a pressure of 1~bar), which was followed by temporal compression using two chirped mirrors (PC70, Ultrafast Innovations), with a GDD of $-40$~fs$^2$ each. These mirrors were placed inside a vacuum chamber, allowing us to fine-tune the chirp by varying the air pressure inside the chamber. A pressure of 0.45~bar was used in the current study. Another spherical mirror with a focal length of 2.5~m was used to image the pulses into a third tube, where a He pressure of 0.6~bar was applied. The final temporal compression was achieved by six PC70 mirrors as well as a pair of wedges. One of these wedges was mounted on a translation stage to fine-adjust the temporal compression.

The spectrum of the incoming laser and the spectra recorded after each compression stage are shown in Fig.~\ref{figure1}(b), demonstrating that an octave-spanning spectrum is generated after the third stage. The central wavelength is blue-shifted, as is typically observed in HCF compression~\cite{nagy21}, which may be attributed to self-steepening~\cite{demartini67}. In addition, while we have taken care to keep ionization at low levels, ionization-induced SPM~\cite{bloembergen73, major23} may contribute to the blue shift observed in the broadened spectrum. The pulse energy after compression was 4.9~mJ, corresponding to a transmission efficiency of 55~$\%$, which is higher than the values obtained from HCFs in a similar parameter regime~\cite{quille20, nagy20}. An excellent single-shot pulse energy stability of 0.17~$\%$ r.m.s. was obtained over half an hour (see Fig.~\ref{figure1}(c)), which compares favorably to the schemes presented in Refs.~\cite{quille20} (0.5~$\%$ over 5 minutes), \cite{toth23} (2.4~$\%$ over 2 hours) and~\cite{rivas17} ($<2$~$\%)$. To obtain the best possible sensitivity from our energy meter in this measurement, an iris was placed in the beam and was partially closed, reducing the pulse energy to 1.88~mJ and giving a sensitivity of 1~$\mu$J. 

Temporal characterization of the compressed pulses was performed using the SEA-F-SPIDER (spatially encoded arrangement for direct electric field reconstruction by spectral shearing interferometry) technique~\cite{witting11}. For these measurements, the NIR beam was clipped by an iris to a diameter of 12~mm, representing the part of the beam that was used for HHG after optimization. The NIR pulse energy after the iris was 3.8~mJ. Fig.~\ref{figure2}(a) depicts the spatio-spectral intensity distribution at the focal plane. As shown in Fig.~\ref{figure2}(b), a good homogeneity of the spatio-temporal intensity distribution was obtained. Following spatial integration, a pulse duration of $3.69 \pm 0.04$~fs was retrieved over ten consecutive measurements (see Fig.~\ref{figure2}(c)), corresponding to $1.54 \pm 0.02$ optical cycles. The quality of the obtained results is comparable to the results retrieved with a hollow-core fiber that was characterized using SEA-F-SPIDER~\cite{witting11}.

\begin{figure}[tb]
 \centering
 \includegraphics[width=0.5\textwidth] {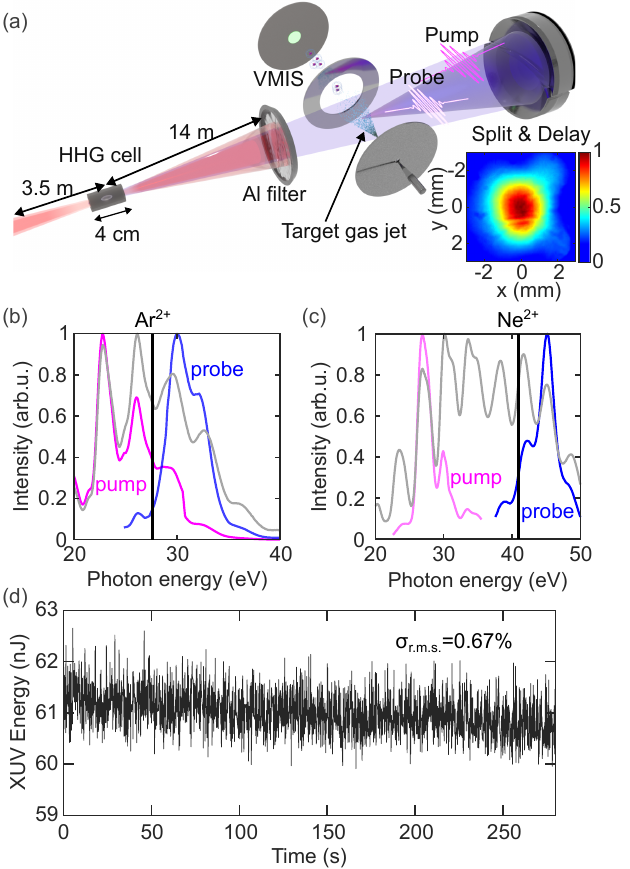}
 \caption{\label{figure3} APAPS setup. (a) Intense attosecond pulses are generated using an 18-m-long HHG beamline. Following attenuation of the NIR driving laser pulses using a 100-nm-thick Al filter, a split-mirror setup is used to select different spectral bands for the pump and the probe pulses. A pulsed gas jet provides Ar or Ne atoms, and the generated ions are detected using a VMIS. The XUV beam profile before focusing is shown in the inset. (b) XUV spectrum obtained for HHG in Xe (gray) as well as the XUV pump (magenta) and probe spectra (blue) used in the Ar experiment. The latter were obtained by convoluting the XUV spectrum with simulated XUV mirror reflectivities. The mirror used for the pump beam was coated with B$_4$C, while the mirror used for the probe beam was coated with Sc/Si. The second ionization potential of Ar is indicated by the black vertical line. (c) Same for the experiment in Ne. Here the mirror used for the pump beam was coated with Sc/Si, while the mirror used for the probe beam was coated with Mo/Si. (d) Single-shot XUV pulse energy stability, giving a value of 0.67~$\%$ r.m.s. The right panel shows the distribution of measured pulse energies.}
\end{figure}

\section{All-attosecond pump-probe spectroscopy}
To demonstrate that the presented pulse compression scheme is ideally suited for APAPS, we have performed ion spectroscopy experiments in Ar and Ne. Intense attosecond pulses were generated using an 18-m-long HHG beamline~\cite{senfftleben20, kretschmar22}, see Fig.~\ref{figure3}(a). The compressed NIR pulses were focused using a telescope consisting of a concave mirror ($f=0.75$~m) and a convex mirror ($f=-1$~m). Harmonics were generated in a 4-cm-long gas cell filled with Xe or Kr, which was placed at a distance of 3.5~m from the last NIR mirror. An iris was used to optimize the HHG yield, resulting in an optimal iris diameter of 12~mm. After HHG, the generated XUV pulses co-propagated with the NIR driving pulses for about 14~m, and a 100-nm-thick Al filter was inserted to block the NIR beam. Two spherical mirrors (25~cm focal length) with different single- or multilayer coatings were used to select different spectral bands for the attosecond pump and probe pulses. The obtained XUV spectra following reflection from the mirrors used in the Ar experiment and in the Ne experiment are shown in Figs.~\ref{figure3}(b) and \ref{figure3}(c), respectively. The temporal delay was controlled using a closed-loop nano-positioning stage for the lower XUV mirror. Ion measurements were performed using a velocity-map imaging spectrometer (VMIS)~\cite{eppink97} operated in spatial-imaging mode~\cite{stei13}. Ar and Ne atoms were provided by a repeller-integrated valve~\cite{ghafur09}.

The XUV pulse energy in Kr was measured after the Al filter using an XUV photodiode (AXUV100G), giving a value of 61~nJ. Taking into account a transmission of the Al filter of about 30~$\%$, this corresponds to an XUV pulse energy of about 200~nJ at the source. The stability of the single-shot XUV pulse energy was measured to be 0.67~$\%$ r.m.s. over about 5 minutes (see Fig.~\ref{figure3}(d)). This is an excellent value taking into account that the carrier-envelope phase (CEP) of the driving laser was not stabilized. Moreover, this value compares favorably to the value of 4.1~$\%$ obtained over 100 shots in Ref.~\cite{xue20}. Assuming that about 50~$\%$ of the XUV pulse energy was incident on each of the XUV half-mirrors, the XUV pump and probe pulse energies on target in the experiment in Ar were measured to be 2.6~nJ and 4.5~nJ, respectively. In the experiment in Ne, the pump and probe pulse energies were 2.9~nJ and 2.2~nJ. Based on a spatial cross-correlation measurement~\cite{kretschmar22, kretschmar24}, the XUV beam waist radii in the horizontal and vertical directions were estimated to be 1.17~$\mu$m and 2.34~$\mu$m, respectively. Together with an estimation of the attosecond pulse structure (see below), the XUV pump and probe peak intensities in the Ar experiment were estimated as $1.5 \times 10^{14}$~W/cm$^2$ and $2.6 \times 10^{14}$~W/cm$^2$, respectively. In the Ne experiment the XUV pump and probe peak intensities were estimated as $1.7 \times 10^{14}$~W/cm$^2$ and $1.3 \times 10^{14}$~W/cm$^2$, respectively.

\begin{figure*}[tb]
 \centering
 \includegraphics[width=0.6\textwidth] {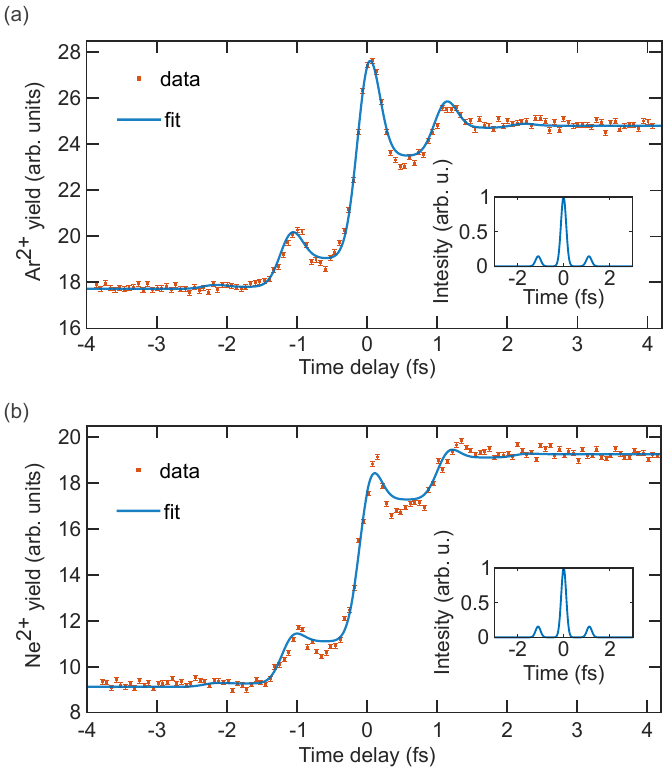}
 \caption{\label{figure4} APAPS results. (a) Ar$^{2+}$ ion yield as a function of the time delay between an attosecond pump pulse and an attosecond probe pulse (see Fig.~\ref{figure3}(b) for the corresponding spectra). The increase of the ion yield when the probe pulse arrives after the pump pulse (positive time delays) is attributed to sequential double ionization of Ar. The oscillations around zero time delay are due to simultaneous two-photon absorption when the pump and the probe pulses overlap in time. The data were averaged over 2~s at each time delay. The error bars are the standard deviation obtained from the fluctuations in a separate measurement where the delay was not varied. The blue curve shows a fit based on a simple model, indicating an attosecond pulse structure as shown in the inset. (b) APAPS in Ne (see Fig.~\ref{figure3}(c) for the corresponding spectra) shows qualitatively similar behavior as in the Ar measurement. The blue curve shows a fit, giving a similar pulse structure as in (a) (see inset).}
\end{figure*}

 The results of the Ar experiment are presented in Fig.~\ref{figure4}(a). The Ar$^{2+}$ ion yield is increased at positive time delays, when the probe pulse follows the pump pulse. In accordance with the results obtained in Ref.~\cite{kretschmar24}, this can be explained by sequential two-photon absorption, where the pump pulse generates Ar$^+$ ions and where the probe pulse further ionizes Ar$^+$, leading to the formation of Ar$^{2+}$. The signal increase for positive time delays is due to the fact that the XUV probe pulse is dominated by spectral contributions lying above the second ionization potential (IP) of Ar at 27.6~eV~\cite{nist}, see Fig.~\ref{figure3}(b). In comparison, the XUV pump pulse spectrum is dominated by contributions below 27.6~eV. In addition, clear oscillations with a period of 1.1~fs are observed around zero time delay in Fig.~\ref{figure4}(a), which are attributed to the simultaneous absorption of two XUV photons when the individual attosecond bursts overlap in time. These oscillations were not clearly observed in our previous experiment, where a HCF was used for spectral broadening and where the XUV intensity was about an order of magnitude smaller~\cite{kretschmar24}. Both the higher stability obtained when using the cascaded post-compression scheme and the higher statistics due to the increased number of two-photon events (as a result of the higher XUV intensity) may contribute to the high quality of the data obtained in the present work. In addition, differences in the XUV spectra that were obtained in Ref.~\cite{kretschmar24} and in the present work, due to the different HHG schemes that were applied, may contribute to the different results. The oscillation period of 1.1~fs is substantially shorter than half the oscillation period of the driving laser (1.3~fs), which is indicative of a transient blueshift of the driving laser in the HHG medium~\cite{kretschmar24}. The data were fitted using a simple model~\cite{kretschmar22, kretschmar24} (blue curve in Fig.~\ref{figure4}(a)) taking into account both sequential and simultaneous two-photon absorption (see Supplemental Material). This fitting suggests an attosecond pulse structure as shown in the inset of Fig.~\ref{figure4}(a), where the FWHM of the individual attosecond bursts is 269~as, and the relative intensities of the pre-and post-pulses with respect to the main pulse are 14~$\%$. We note that the obtained results are averaged with respect to the CEP of the driving laser. In the future, the time resolution may be further improved by single-shot CEP tagging in combination with single-shot recording of the experimental data. Alternatively, a CEP-stable laser system may be used. The total scanning time of the measurement shown in Fig.~4(a) was four minutes, which compares favorably to the measurement time of more than three hours of the first APAPS experiment using near-isolated attosecond pulses~\cite{takahashi13}. At the same time, the data quality obtained in the present study is at a substantially higher level.
 
To be able to access inner-shell excitations, it is desirable to develop APAPS capabilities at higher photon energies. As a step towards this goal, we have performed APAPS in Ne. The second ionization potential of Ne is at 41.0~eV~\cite{nist}. Compared to previous APAPS experiments performed in Xe~\cite{tzallas11}, N$_2$~\cite{takahashi13} and Ar~\cite{kretschmar22, kretschmar24}, APAPS in Ne is substantially more demanding because of the decrease of the HHG conversion efficiency towards higher photon energies and because the ionization cross sections of Ne are substantially smaller~\cite{chan92, chan92b, chan93}. The result of this measurement is shown in Fig.~\ref{figure4}(b) and qualitatively resembles the measurement in Ar. The increase of the Ne$^{2+}$ ion yield is explained by the fact that only the XUV probe pulse has significant spectral distributions lying above the second ionization of Ne, see Fig.~\ref{figure3}(c). A fit of these data (blue curve) results in a similar pulse structure as obtained for Ar (see inset of Fig.~\ref{figure4}(b)). The width of the individual attosecond bursts in this case was 256~as and the relative intensities of the pre- and post-pulses with respect to the main attosecond burst was 15~$\%$.

\section{Conclusions and outlook}
In conclusion, we have demonstrated a simple scheme for the compression of multi-mJ pulses down to 1.5 optical cycles. The high stability of the pulse duration and the pulse energy makes this approach ideal for the generation of intense attosecond pulses, which in our laboratory are now available as a turn-key system that is operational within a few minutes. The development of the pulse compression scheme has allowed us to perform APAPS in Ar and Ne with an unprecedented quality of the obtained data. This paves the way to performing even more demanding experiments such as all-attosecond transient absorption spectroscopy (AATAS). In the future, our approach may be scaled both to lower pulse energies (to reduce the required space), and to higher pulse energies and multi-TW powers. The required space in the latter case is compatible with the space that is available at large-scale facilities~\cite{kuhn17, hort19} and in university laboratories~\cite{rudawski13, wang18} where suitable multi-TW lasers are available. 

\section*{Acknowledgments}
We thank M. Krause, C. Reiter and R. Peslin for their technical support. Furthermore, we thank the Fraunhofer Institute for Applied Optics and Precision Engineering IOF in Jena for providing the simulated reflectivity curves for the XUV multilayer mirrors.

\section*{Disclosures}
The authors declare no conflicts of interest.

\section*{Data availability statement}
Data underlying the results presented in this paper are not publicly available at this time but may be obtained from the authors upon reasonable request.

\section*{Supplemental document}
See Supplement 1 for supporting content.

\bibliography{Bibliography}

\end{document}